\documentclass[intlimits,twoside,a4paper]{article}

\usepackage{amsmath,amssymb}
\usepackage{graphicx}

\usepackage[T2A]{fontenc}
\usepackage[cp1251]{inputenc}

\usepackage[eqsecnum]{cmpj3}

\issue{2019}{22}{3}{33001}
\doinumber{10.5488/CMP.22.33001}

\title[Discriminate the quality of pear fruit using silicene nanosheet]%
{Silicene nanosheet to discriminate the quality of pear fruit based on volatiles adsorption --- a DFT application }%
\author[R. Keerthi Bhavadharani, V. Nagarajan, R. Chandiramouli]{R. Keerthi Bhavadharani\refaddr{label1}, V. Nagarajan\refaddr{label2}, R. Chandiramouli\refaddr{label2}\thanks{Corresponding author} }
\addresses{
\addr{label1} School of Chemical and Biotechnology, SASTRA Deemed University, \\Tirumalaisamudram, Thanjavur --- 613 401, India 
\addr{label2} School of Electrical and Electronics Engineering, SASTRA Deemed University,\\ Tirumalaisamudram, Thanjavur --- 613 401, India 
}

\date{Received February 28, 2019, in final form June 20, 2019}

\begin{document}

\maketitle

\begin{abstract}
We report the interaction between the silicene nanosheet (Si-NS) and volatile organic compounds (VOCs) released from the pear fruit (Pyrus communis) in ripened and over-ripened stages using density functional theory (DFT) technique. The geometric stability of Si-NS is studied from the phonon band structure. Further, the electronic property of Si-NS is studied from the energy band gap structure, and the energy gap is found to be 0.46~eV, which exhibits semiconductor property. The outcomes infer that the adsorption of volatiles released from the pear fruit on silicene nanosheet is in the following order hexyl acetate $\rightarrow$ butyl acetate $\rightarrow$ butyl butyrate in the ripened stage whereas in the over-ripened stage the adsorption sequence is noticed to be acetic acid $\rightarrow$ ethyl acetate $\rightarrow$ 1-butanol. The adsorption property of pear fruit volatiles on silicene nanosheet is documented with the adsorption energy, average energy gap changes, and Bader charge transfer. Moreover, the adsorption of VOCs on silicene nanosheet is also explored using the energy band structure, electron density along with the adsorption sites and density of states (DOS) spectrum. Besides, the findings reveal that the silicene nanosheet can be used to discriminate the quality of pear fruit.

\pacs 07.07.Df, 61.46.+w
\end{abstract}

\section{Introduction}

In recent days, significant work is carried out in two-dimensional (2D) materials, namely graphene and its cousins, transition metal dichalcogenides, MXenes, and so on. Owing to the tunable chemical and electronic properties of 2D materials, researchers focus on the cutting-edge application in engineering and in biological fields \cite{1,2,3,4}. The class of 2D materials possessed layered structures with strong chemical bonds along its plane while weak chemical bonds in the out of plane link are noticed in these nanostructures. Thus, the monolayer or few-layers can be sliced into individual separate atomic layers. Moreover, cleaving the bulk material into monolayer or few-layer results in the extraordinary variations in their electronic properties due to the quantum confinement effect. Thus, the varying electronic and chemical properties of 2D materials lead the scientific community to think over the novel applications. One of the 2D layered materials is silicene. Similar to graphene, silicene is Si analogous and is a newly synthesized 2D material with exceptional features and very promising applications \cite{5,6}. Furthermore, silicene material is highly reactive rather than flat graphene structure. Hence, functionalization of silicene plays an important role in modifying its electronic properties, which may be suitable for  various micro and nanoelectronic devices.  Besides, the DFT method clearly predicts that 2D honeycomb monolayer of Si atoms is highly stable and exhibits buckled structure due to its $sp^3$ hybridization, in contrast to flat graphene structure. However, the edge border termination and functionalization of silicene layers with hydrogen and modification of the surface with unsaturated organic molecules lead to novel applications of silicene \cite{7}. Moreover, silicene exhibits mixed $sp^2$ and $sp^3$ hybridization of silicon atoms in silicene sheets.  Drummond et al. reported about the electronic structure of silicene under the external field, which is oriented normal to the silicene sheet. The authors report that the electric field can be employed to fine-tune the band gap \cite{8}.
  	
It is known that the volatile organic compounds (VOCs) are generally present whenever we sense the scent from the source. Moreover, VOC emission from the plant and trees fruits and flowers attracts the animals, which leads to the dispersal of seeds to other places after consumption by animals. Nevertheless, VOCs comes out from the various periods of fruits, which infer its quality and freshness \cite{9}. The general VOC compounds of pear fruit (\textit{Pyrus communis}) present in the ripen stage are butyl acetate, butyl butyrate, hexyl acetate whereas in the over-ripen stage they will be converted into 1-butanol, acetic acid, and ethyl acetate, respectively. Besides, esters have pleasing odours, and certainly, they are the major components of odour and flavour for several fruits \cite{10}. Besides, different fruits emit aroma, which is distinct from other fruits. The aroma emitted from the fruits depends upon the combination of various volatiles. Moreover, the aroma emission by a particular fruit change due to aging, in which the volatiles turn to other combinations due to microbial action on the fruit. Moreover, the aroma of the fruit is owing to different compounds such as lipid-derived compounds, mono and sesquiterpenes, phenolic derivatives and amino-acid based compounds \cite{11}. We have reported the quality estimation of banana fruit based on the volatiles evolved during the ripened and over-ripened state \cite{12, 13}.  The novel piece of the article is to afford the adsorption property of VOCs of pear fruits, namely butyl acetate, butyl butyrate, hexyl acetate, 1-butanol, acetic acid and ethyl acetate in silicene nanosheet, which is examined and reported.

\begin{figure}[!b]
\vspace{-3mm}
\centerline{\includegraphics[width=0.6\textwidth]{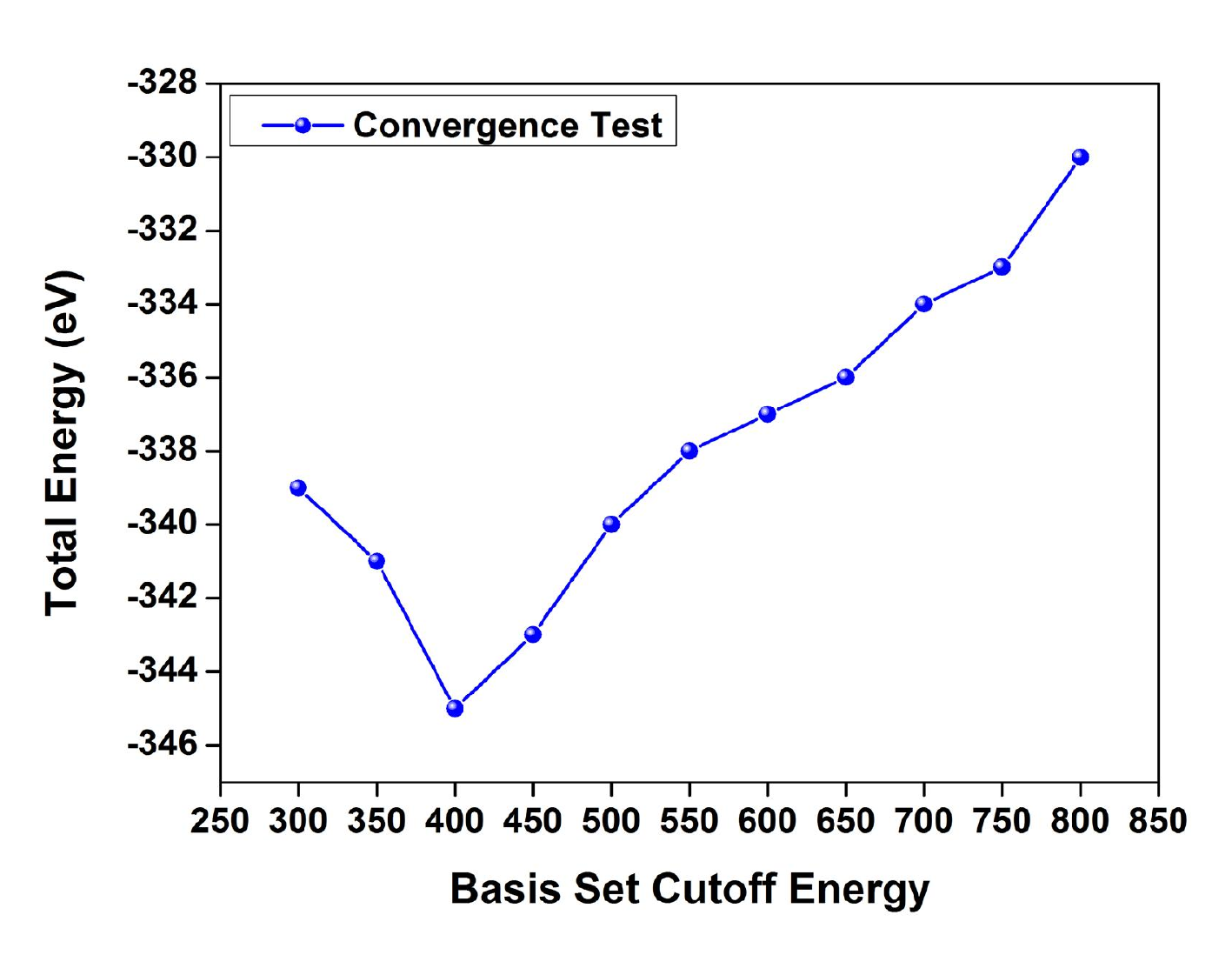}}
\caption{(Colour online) Convergence test for pristine silicene nanosheet.} \label{fig-s1}
\end{figure}

\section {Quantum chemical calculations}
The geometric structure of silicene nanosheet and the four vapors emitted from the pear fruit are adsorbed on pristine silicene sheet, which is studied by DFT method as implemented in the SIESTA code \cite{14}. In order to describe the weak dispersive interactions, we used van der Waals dispersion correction (vdW-DF) in accordance with GGA/PBE exchange-correlation functional as implemented in DFT method \cite{15, 16}. We kept the grid mesh cut-off as 400 eV during geometrical optimization. The convergence test for pristine silicene nanosheet is shown in figure~\ref{fig-s1}.  Besides, the Brillouin zone integration of silicene sheets is sampled with ($15 \times 15 \times 1$) gamma centered Monkhorst-Pack $k$-grid \cite{17}. We calculated electron density, energy band gap structure, and DOS-spectrum of Si-NS using the SIESTA code. The electronic transfer between the target vapors and silicene sheets is studied using Bader-atoms-in molecules (BAIM) exploration \cite{18}. In addition, the full geometric relaxation is performed using the conjugate-gradient algorithm until the Hellmann-Feynman force was congregated to 0.01~eV/\AA.

In the proposed work, the double-zeta polarization (DZP) basis sets were utilized to study the electronic properties of silicene nanosheet with norm-conserving Troullier Martins pseudopotentials \cite{19, 20}. Further, the vacuum padding of 16~{\AA} along $X$ and $Y$ axes was maintained in order to eliminate the interaction among the neighboring layers.

\section {Results and discussion}

\subsection {Silicene nanosheet structure and its electronic properties}

\begin{figure}[!b]
\centerline{\includegraphics[width=0.6\textwidth]{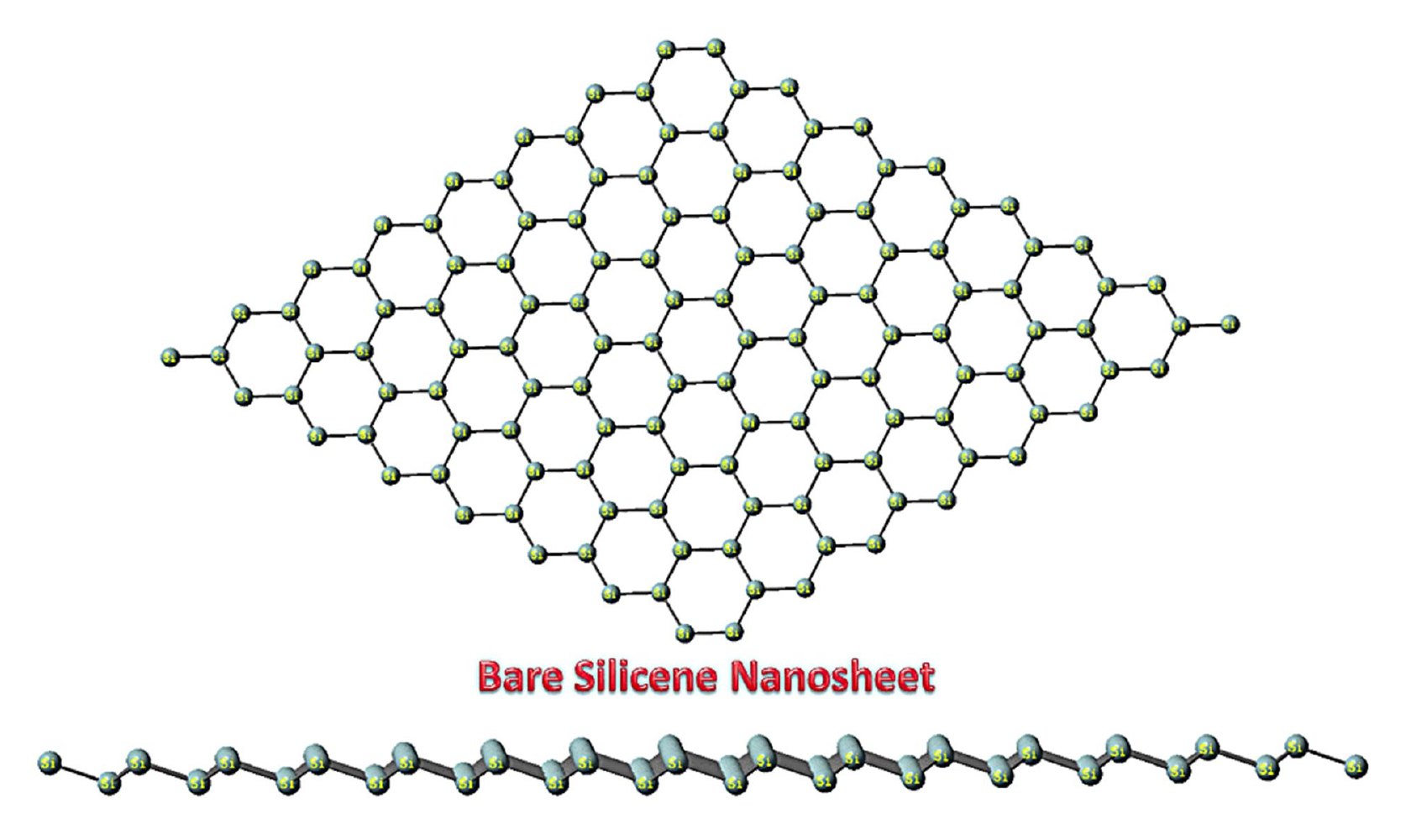}}
\caption{(Colour online) Representation of silicene nanosheet.} \label{fig-s2}
\end{figure}

The Si-NS material is utilized as a noteworthy substrate for the adsorption of VOCs emanated from the pear fruit. Moreover, due to the $sp^3$ hybridization of silicene sheet, the bucking is noticed in its nanostructure as presented in figure~\ref{fig-s2}. The supercell size of silicene nanosheet is $8 \times 8 \times 1$.

\begin{figure}[!b]
\centerline{\includegraphics[width=0.60\textwidth]{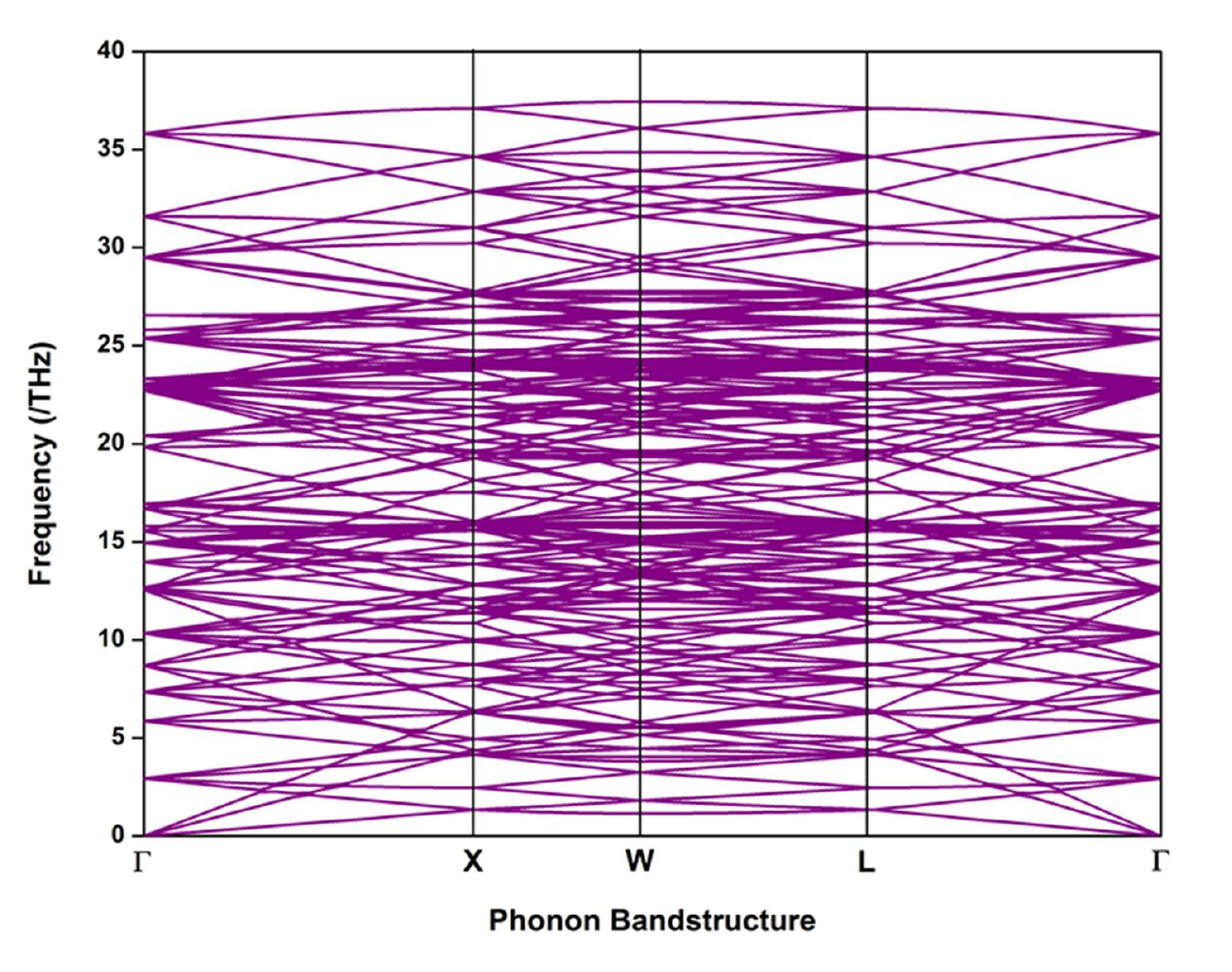}}
\caption{(Colour online) Phonon band structure of pristine silicene nanosheet.} \label{fig-s3}
\end{figure}

Moreover, the stable structure of silicene nanosheet is established using the phonon band structures, which is depicted in figure~\ref{fig-s3}. Nevertheless, no unreal frequency is noticed in the phonon band structure that clearly specifies the stable structure of silicene \cite{21}. The lattice constant, buckling height and bond distance between Si-Si is calculated to be 3.87 \AA, 0.44~{\AA} and 2.26~{\AA}, respectively.

\begin{figure}[!t]
\centerline{\includegraphics[width=0.85\textwidth]{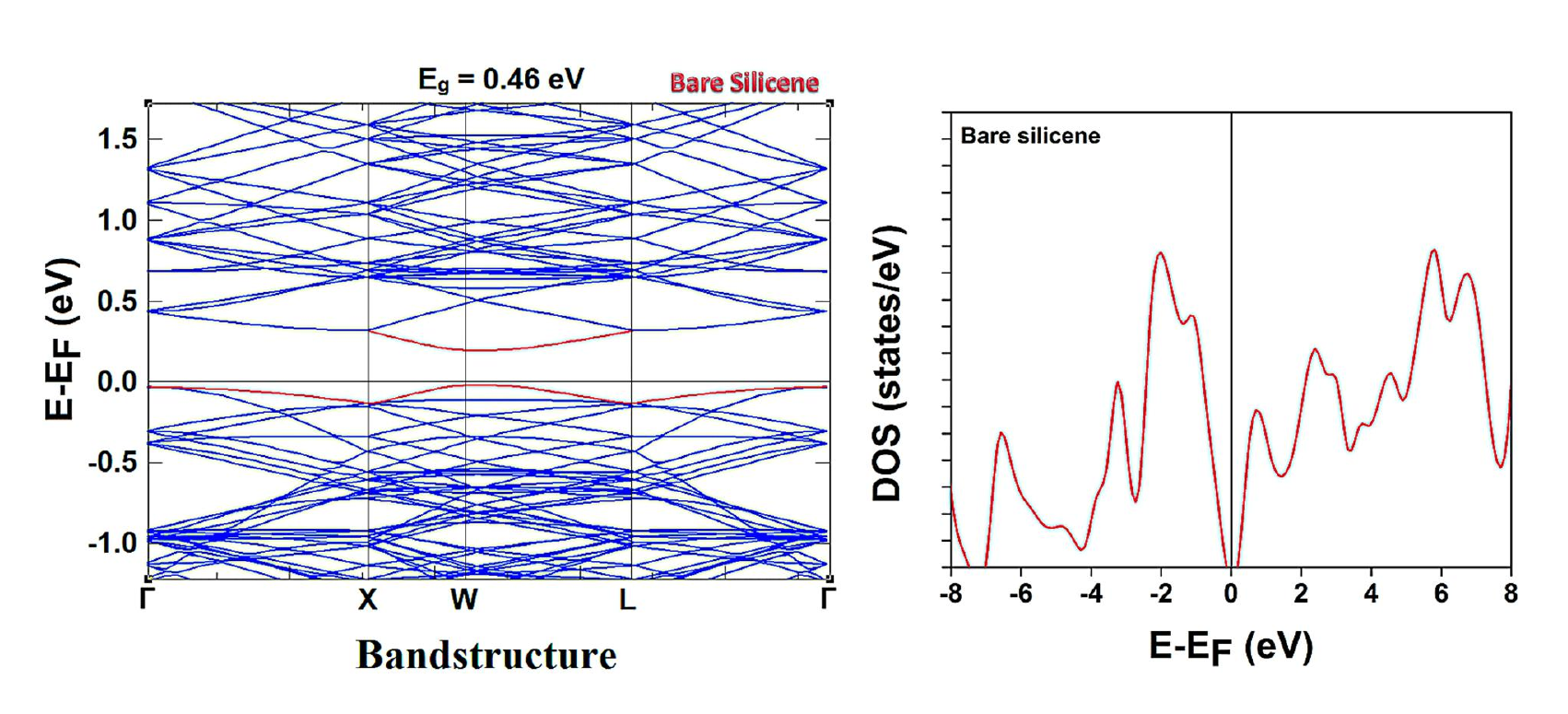}}
\caption{(Colour online) DOS and band structure diagram of bare silicene nanosheet.} \label{fig-s4}
\end{figure}

The energy gap of the zigzag border of silicene nanosheet is found to be 0.46 eV. Besides, the peak maxima are also noticed in  various energy intervals both in the highest occupied (HO-) valence band and in the lowest unoccupied (LU-) conduction band as perceived in the DOS spectrum. The energy gap of 0.46 eV favours the use of silicene sheet as a base substrate for chemo biosensor \cite{22}. Figure~\ref{fig-s4} exemplifies the band gap structure and DOS-spectrum of silicene nanosheets. The pear fruit quality is assessed by studying the adsorption energy, Bader charge transfer, and the energy gap changes upon the interaction of VOCs on silicene sheets.

\subsection{Adsorption of VOCs emitted from pear fruit on silicene nanosheet surfaces}

\begin{figure}[!t]
\centerline{\includegraphics[width=0.65\textwidth]{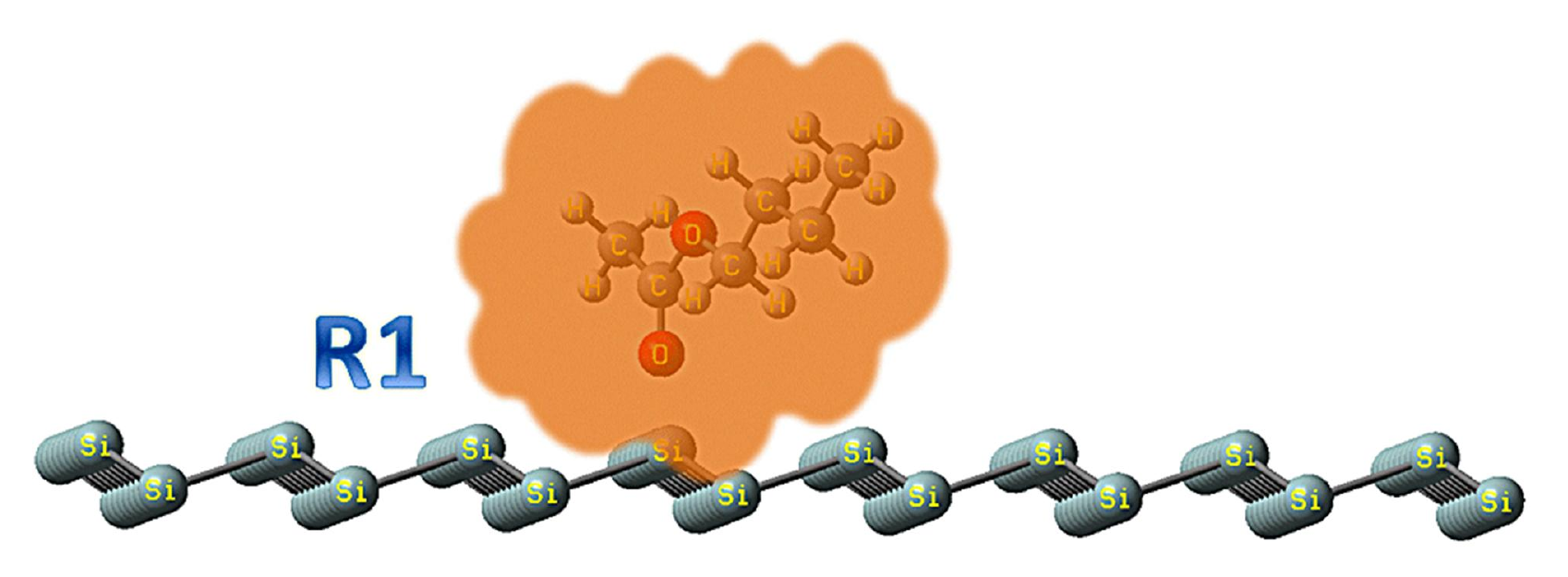}}
\caption{(Colour online) Position R1 --- adsorption of butyl acetate on BSi-NS.} \label{fig-s5}
\end{figure}

\begin{figure}[!t]
\centerline{\includegraphics[width=0.65\textwidth]{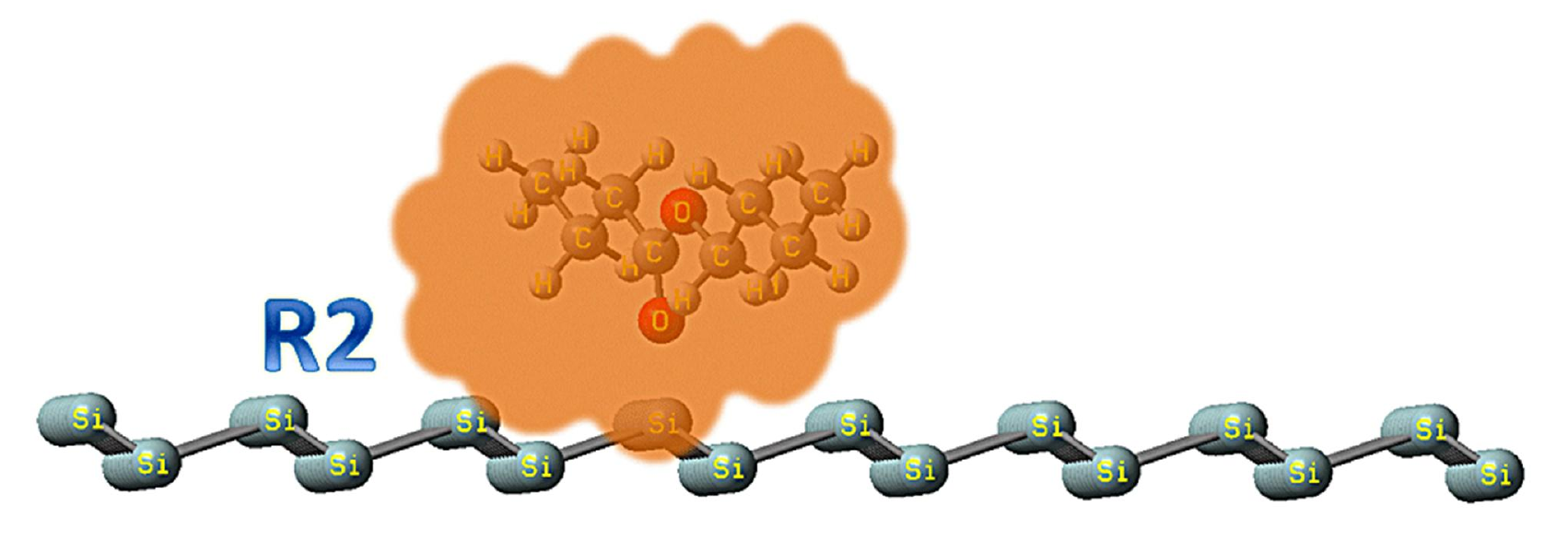}}
\caption{(Colour online) Position R2 --- adsorption of butyl butyrate on BSi-NS.} \label{fig-s6}
\end{figure}

\begin{figure}[!t]
\centerline{\includegraphics[width=0.65\textwidth]{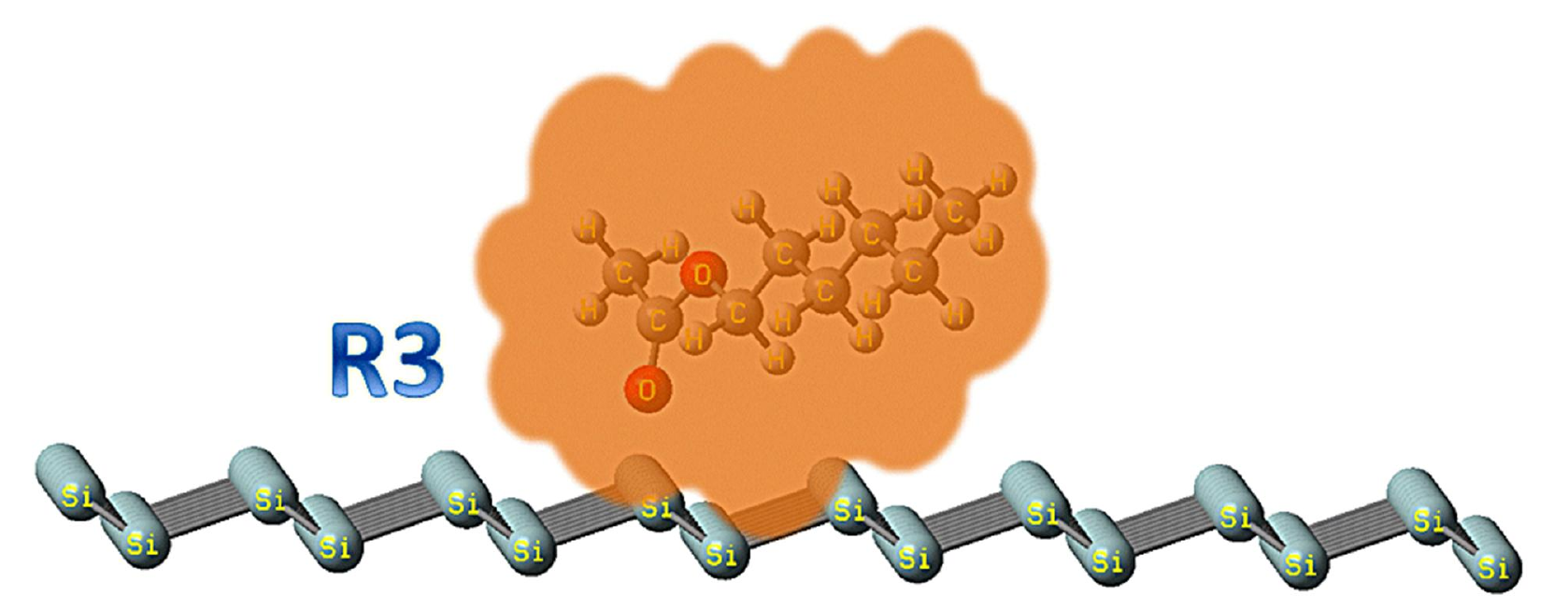}}
\caption{(Colour online) Position R3 --- adsorption of hexyl acetate on BSi-NS.} \label{fig-s7}
\end{figure}

\begin{figure}[!t]
\centerline{\includegraphics[width=0.65\textwidth]{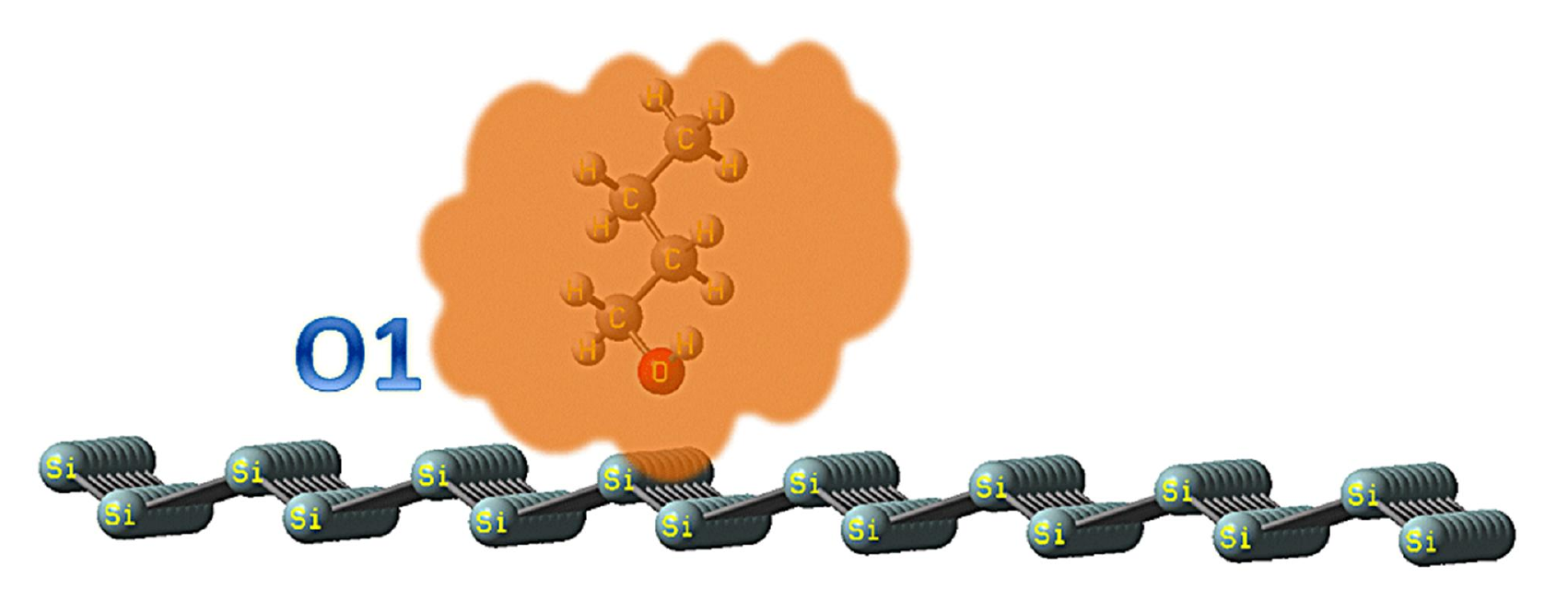}}
\caption{(Colour online) Position O1 --- adsorption of 1-butanol on BSi-NS.} \label{fig-s8}
\end{figure}

\begin{figure}[!t]
\centerline{\includegraphics[width=0.65\textwidth]{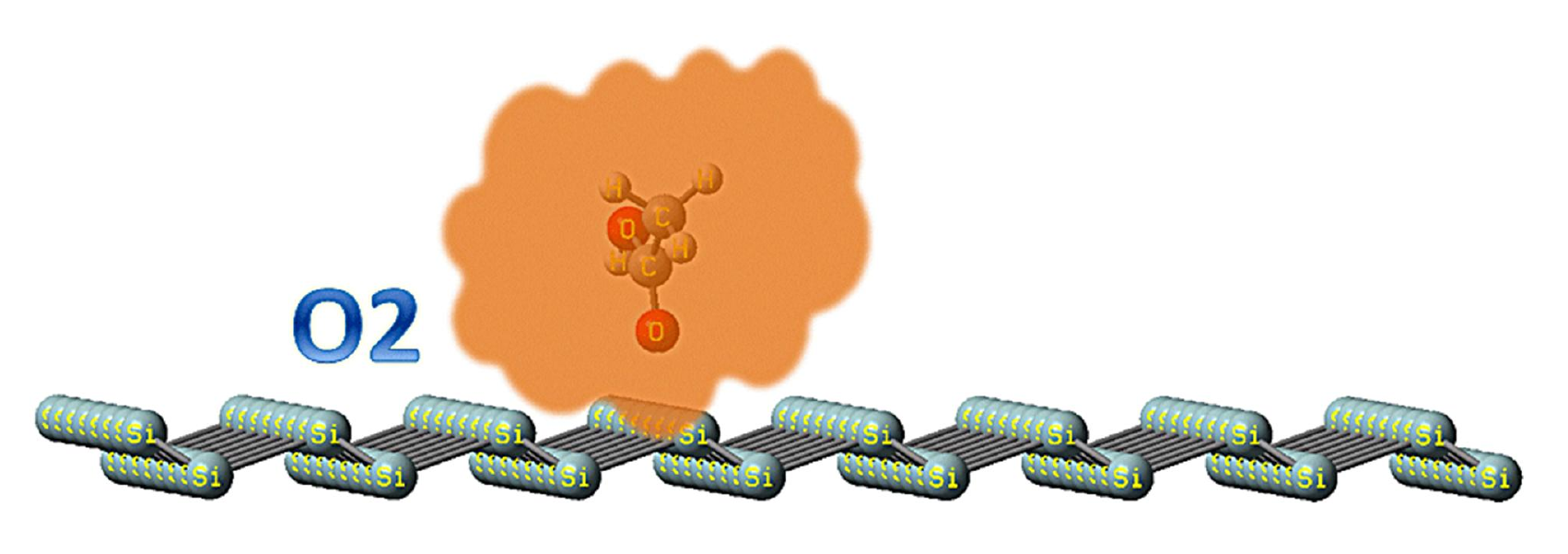}}
\caption{(Colour online) Position O2 --- adsorption of acetic acid on BSi-NS.} \label{fig-s9}
\end{figure}

\begin{figure}[!t]
\centerline{\includegraphics[width=0.65\textwidth]{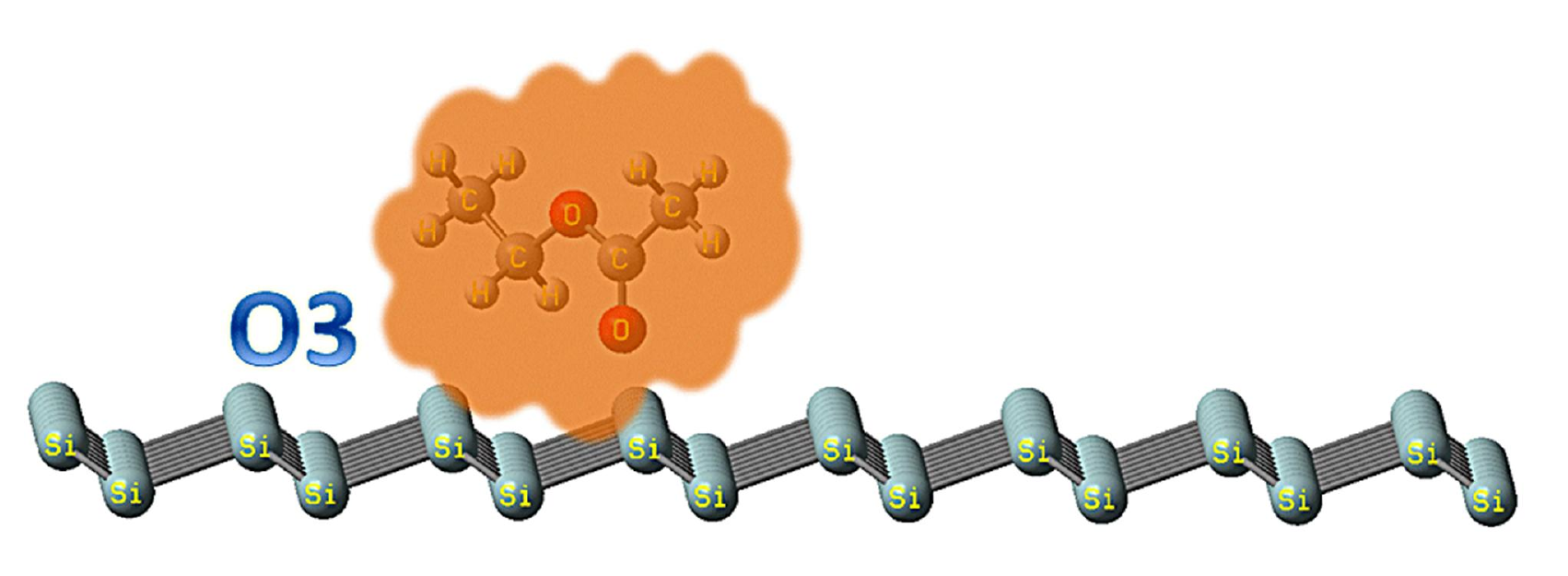}}
\caption{(Colour online) Position O3 --- adsorption of ethyl acetate on BSi-NS.} \label{fig-s10}
\end{figure}

At this juncture, we have to mention and justify the importance of the work. Moreover, it is well-known that at various stages of fruits, different volatile organic compounds (VOCs) are emitted such as esters, mono, and sesquiterpenes, lipid-derived compounds, phenolic derivatives, and amino-acid based compounds, alcohols, etc. Besides, by our olfactory sensor (nose), we can discriminate the ripening stages. Furthermore, with the help of a chemical sensor (electronic nose), it is possible to detect the presence of various VOCs present in the vicinity \cite{23,24,25,26,27,28,29}. The metal oxide sensors are widely used to detect the presence of particular VOC. In recent days, researchers are involved with two-dimensional and one-dimensional materials as chemical sensors to name a few doped graphenes, carbon nanotubes, etc. However, the exfoliation of group IVA, VA materials to a monolayer or few-layer materials leads to the change of its electronic properties. In the present report, we used one of the group IVA materials, silicene sheets exfoliated from bulk silicon as the base material. We begin the adsorption studies of VOCs of pear fruit such as butyl acetate, butyl butyrate, hexyl acetate, 1-butanol, acetic acid and ethyl acetate on silicene nanosheets using GGA/PBE $+$ vdW scheme. Initially, the silicene nanosheet is optimized, and the electronic property exhibits a semiconductor nature. Moreover, a semiconductor nature is a favourable condition for the adsorption of VOCs of pear fruit. Figures~\ref{fig-s5}--\ref{fig-s7} demonstrates the interaction of O atom in the ripened stage of aromas, namely butyl acetate, butyl butyrate and hexyl acetate interacted on Si atom in Si-NS and it is mentioned as position R1, R2, and R3, respectively. Similarly, the interaction of the oxygen atom in over-ripened stage aromas such as 1-butanol, acetic acid and ethyl acetate interacted on Si atom in Si-NS material is referred to as position O1, O2, and O3, respectively, as displayed in figures~\ref{fig-s8}--\ref{fig-s10}.

The adsorption energy of volatiles emitted from the pear fruit upon adsorption on the BSi sheet is studied using adsorption energy ($E_\text{ad}$) \cite{30,31,32,33}, which is given by the equation as follows:
\begin{equation}
E_\text{ad} = [E_\text{silicene/VOC} - E_\text{silicene} - E_\text{VOC} + E_\text{BSSE}],	
\end{equation}
where $E_\text{silicene/VOC}$ represents the adsorption energy of silicene/VOC complex, $E_\text{silicene}$ exemplifies the isolated energy of silicene nanosheet, and $E_\text{VOC}$ depicts the energy of VOCs of pear fruit such as butyl acetate, butyl butyrate, hexyl acetate,1-butanol, acetic acid, and ethyl acetate. $E_\text{BSSE}$ refers to the basis-set superposition-error and has been included in the calculation for neglecting the overlap effects on the basis sets using counterpoise technique. Table~\ref{table1} represents the adsorption energy, average energy band gap changes ($E_\text{g}^\text{a}$) and the energy band gap of butyl butyrate, butyl acetate, hexyl acetate, 1-butanol, acetic acid and ethyl acetate on silicene nanosheet. The adsorption energy of volatiles emitted from pear fruit found in ripened stage such as butyl acetate, butyl butyrate, and hexyl acetate is found to be of higher magnitude.

\begin{table}[!t]
  \renewcommand{\arraystretch} {1.3}
  \caption{VOC adsorption on silicene nanosheet, adsorption energy ($E_\text{ad}$), charge transfer ($Q$), energy gap~($E_\text{g}$) and average energy gap variation ($E_\text{g}^\text{a}$).}
  \label{table1}
  \vspace{2ex}
  \centering
  \begin{tabular}{|c|c|c|c|c|}
  \hline\hline
  Nanostructure 	& $E_\text{ad}$ (eV) 	&$Q$ ($e$)	&$E_\text{g}$ (eV)		&$E_\text{g}^\text{a}~\%$ \\
	\hline\hline
	
BSi-NS	&$-$	&$-$	&0.46		&$-$	\\

\hline
Pear fruit --- Ripened stage  & & & &\\
\hline
Position R1	&$-$1.236	& 0.488 & 0.41	& 10.87 \\
\hline
Position R2	&$-$1.501	& 0.365 & 0.43	& 6.52	\\
\hline
Position R3	&$-$1.561	& 0.451 & 0.40	& 13.04	 \\
\hline
Pear fruit --- Over-ripened stage & & & &\\
\hline
Position O1	&$-$0.927	& 0.403 & 0.38	& 17.39	 \\
\hline
Position O2	&$-$0.577	& 0.531 & 0.35	& 23.91	 \\
\hline
Position O3	&$-$0.803	& 0.480 & 0.36	& 21.74	 \\
\hline\hline
  \end{tabular}
  \end{table}

The adsorption energy of butyl acetate is calculated to be $-$1.23 eV, for hexyl acetate the adsorption energy value is $-$1.56 eV, and for butyl butyrate, $E_\text{ad}$ is calculated as $-1.50$~eV. The higher magnitude of adsorption energy for VOCs found in the ripened stage is administered by the fact of the existence of the ester functional group on all the volatiles. By contrast, the adsorption energy of volatiles emitted in the over-ripened stage has moderate values owing to the presence of carboxylic acid and alcohol group. Moreover, all the VOCs of pear fruit exhibit a negative scale of adsorption energy, which provisions the adsorption of VOC molecules on to the Si-NS. The Bader charge transfer ($Q$) represents the transfer of charge between VOCs of pear fruit with silicene base material \cite{34,35,36,37,38,39}. The sign of charge transfer infers the direction of charge transfer. Besides, the positive magnitude of charge transfer of $Q$ reveals that the charge gets transferred from pear fruit volatiles towards Si-NS. The magnitude of the charge is found to be high for positions R1 and O2, which depends upon the adsorption site. Thus, the charge transfer is also reflected by  various peak maxima in DOS spectrum. Moreover, the positive magnitude of $Q$ is noticed upon adsorption of VOC molecules with Si-NS sheet. The positive value of $Q$ from all the compounds to silicene nanosheet is advocated by the fact that the binding site of volatiles of pear fruit with the silicene nanosheet all have strong electronegative atom such as oxygen. Thus, the transfer of charge takes place from the VOCs on to the Si-NS. Moreover, the changes in the valence band (HOMO) and in the conduction band (LUMO) of both volatiles and Si-NS lead to the relocation of the charge between the VOC and BSi. Thus, it is evident that the change in $Q$ leads to a conclusion that Si-NS behaves as chemiresistor. The energy band gap upon adsorption of pear fruit volatiles with BSi plays a significant part in the interaction behaviour \cite{40,41,42,43,44,45}. The energy gap dissimilarity is observed to be high for VOCs of the over-ripened stage. Why is the bandgap variation high for VOC compounds of pear fruit present in over-ripened stage adsorbed on Si-NS? The reason behind the deviation in the band gap of silicene is governed owing to the interaction of alcohol, carboxylic acid, or ester functional group. Besides, due to the adsorption of VOCs, the band structure gets modified, which in turn varies the band structure of silicene nanosheet. The other significant parameter to infer the interaction of VOCs on silicene nanosheet is expressed by the average energy gap variation ($E_\text{g}^\text{a}$\%) \cite{46}. The $E_\text{g}^\text{a}$  variation is noticed to be high for VOCs present in the over-ripened stage. Moreover, the Si-NS upon adsorption of pear fruit VOCs results in the variation in the resistance. Besides, the change in the average energy gap variation for the over-ripened VOCs of pear fruit is found to be significant rather than the average energy gap variation of the ripened stage of pear fruit. The overall results support that the interaction of VOCs released from the pear fruit leads to modification in the resistance. Hence, depending upon the adsorption of VOCs, it leads to a variation in the resistance; a two-probe silicene nanosheet device can be used to perceive the quality of pear fruit. The change in the current from the two-probe silicene nanosheet device leads to the detection of VOCs.

\subsection {Electronic structure of silicene nanosheet-VOCs of pear fruit}

\begin{figure}[!b]
\vspace{-3mm}
\centerline{\includegraphics[width=0.85\textwidth]{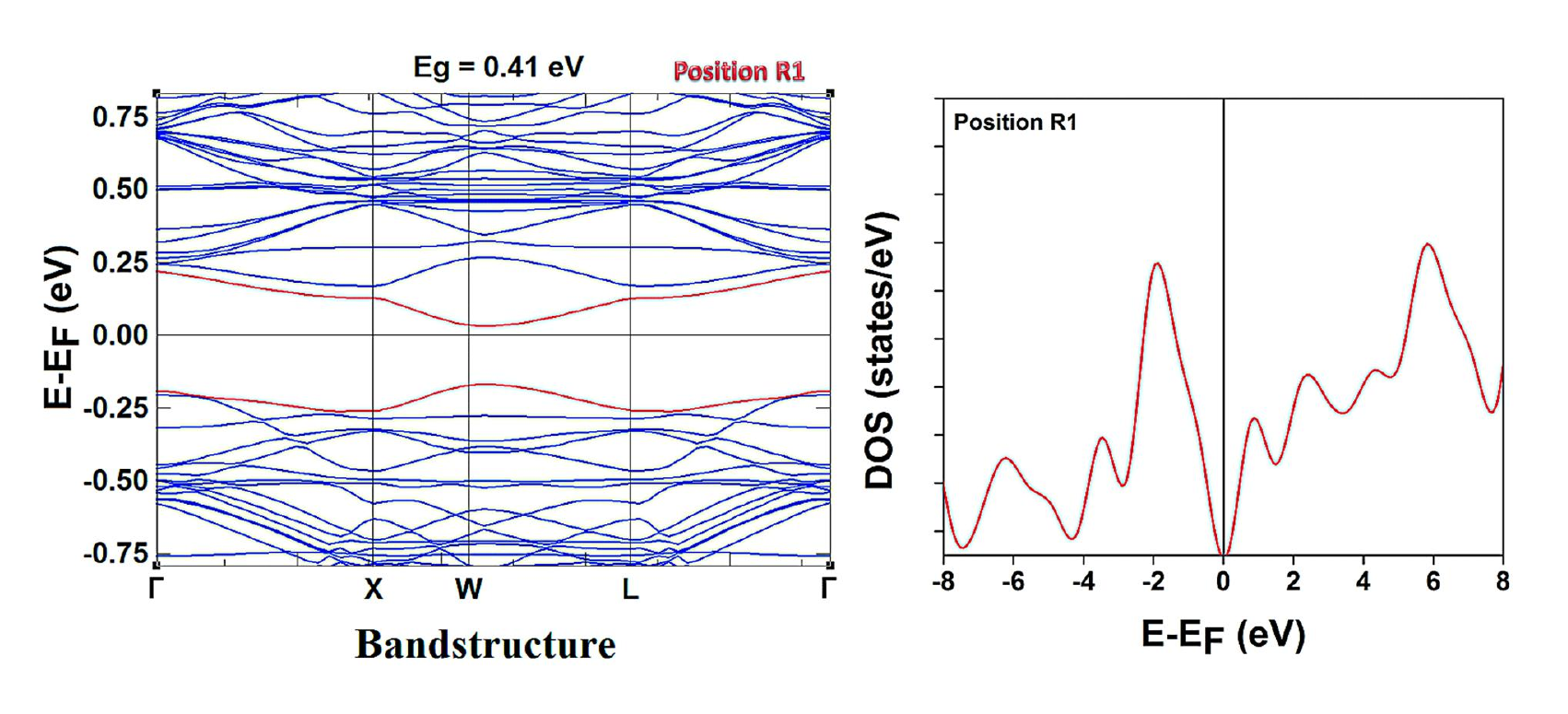}}
\caption{(Colour online) DOS and band structure diagram for position R1.} \label{fig-s11}
\end{figure}

\begin{figure}[!t]
\centerline{\includegraphics[width=0.85\textwidth]{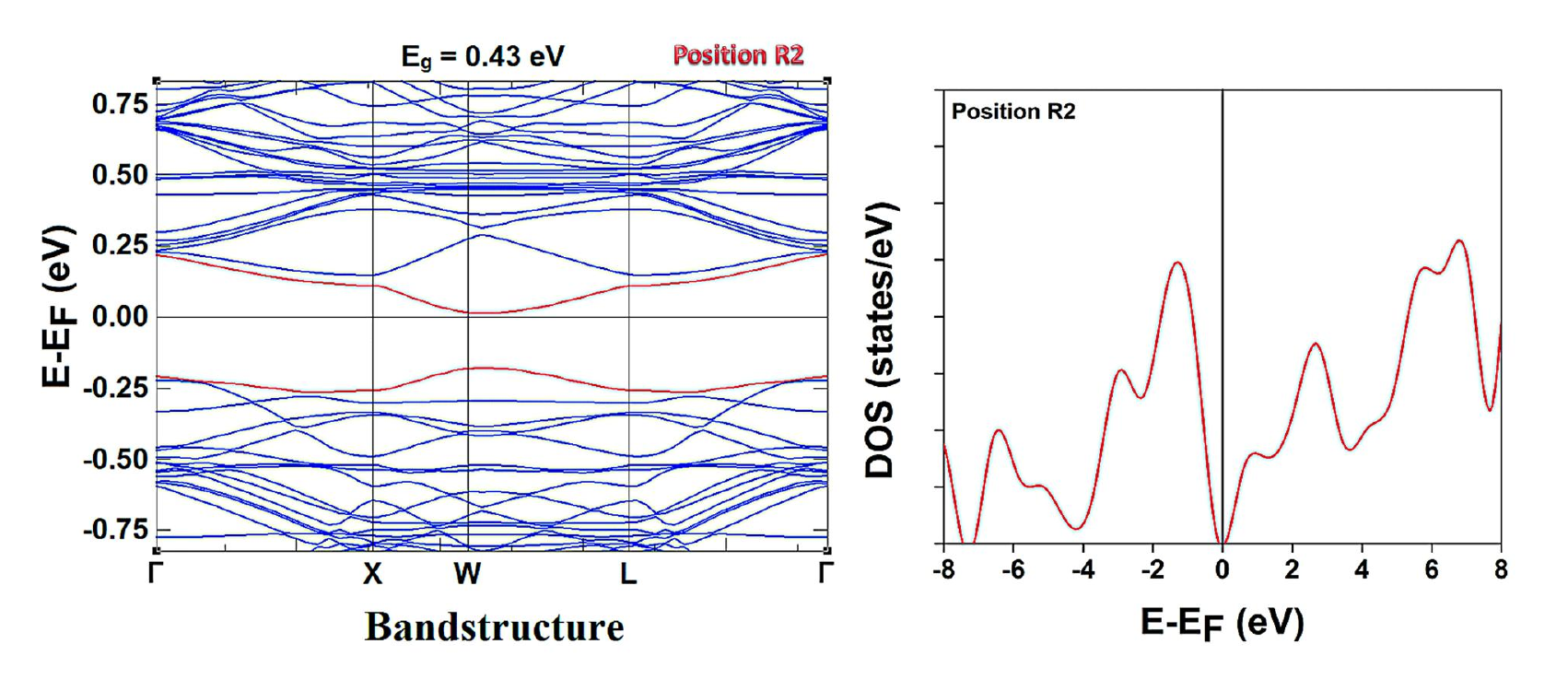}}
\caption{(Colour online) DOS and band structure diagram for position R2.} \label{fig-s12}
\end{figure}

\begin{figure}[!t]
\centerline{\includegraphics[width=0.85\textwidth]{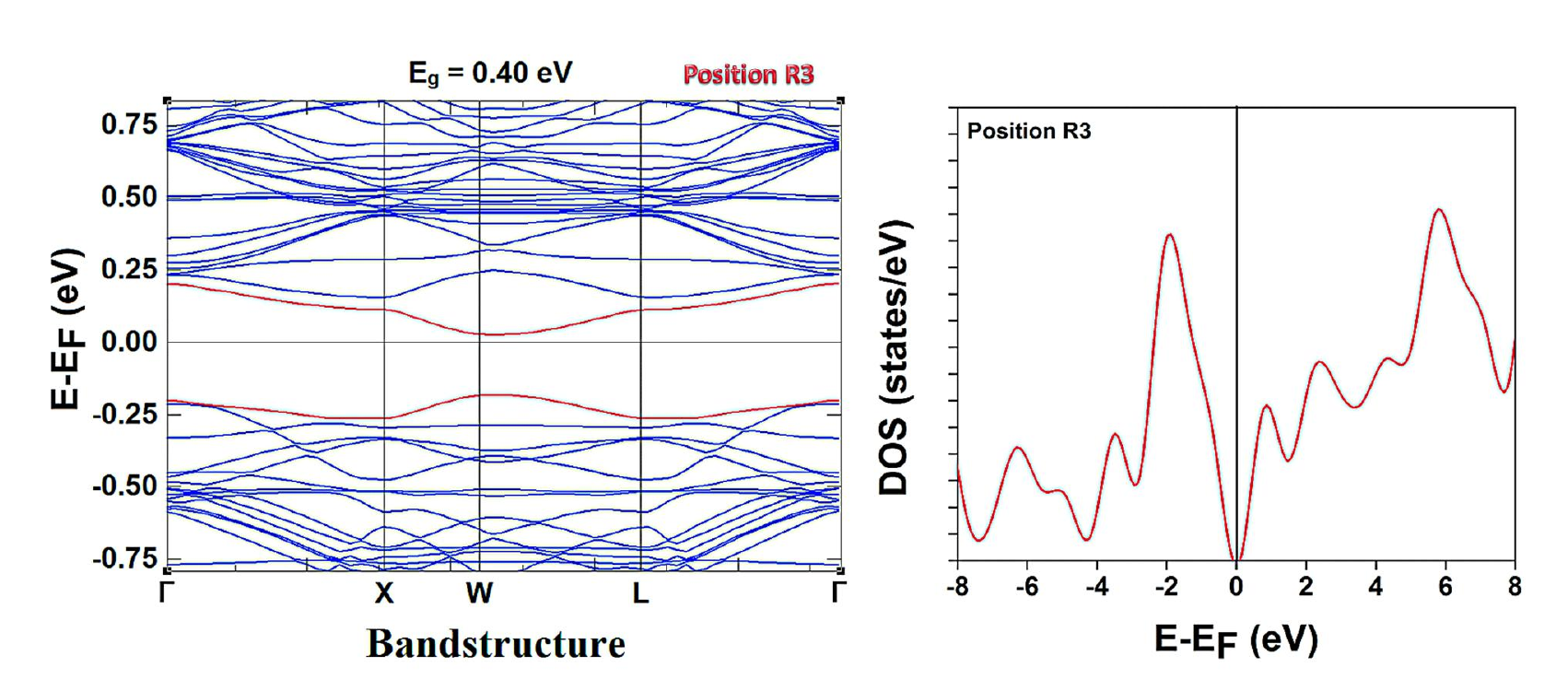}}
\caption{(Colour online) DOS and band structure diagram for position R3.} \label{fig-s13}
\end{figure}

\begin{figure}[!t]
\centerline{\includegraphics[width=0.85\textwidth]{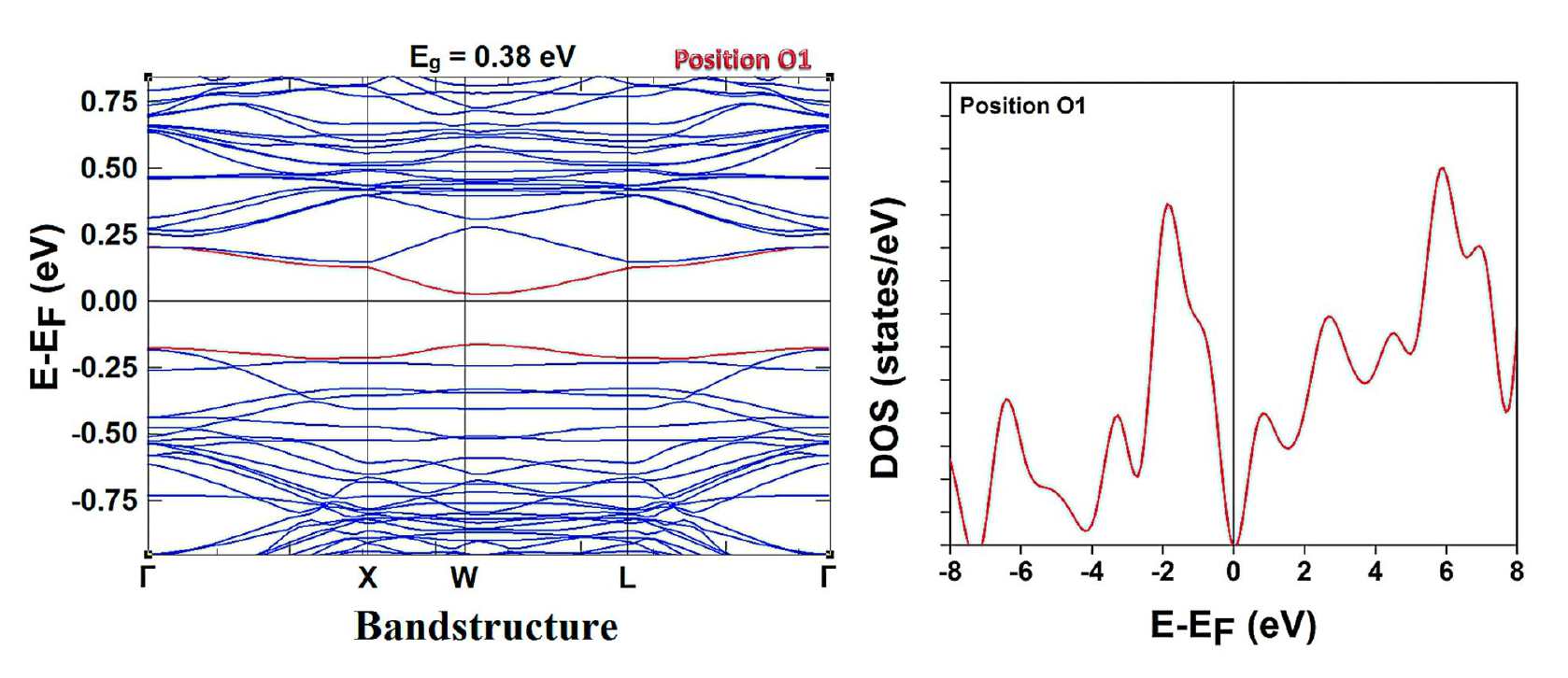}}
\caption{(Colour online) DOS and band structure diagram for position O1.} \label{fig-s14}
\end{figure}

\begin{figure}[!t]
\centerline{\includegraphics[width=0.85\textwidth]{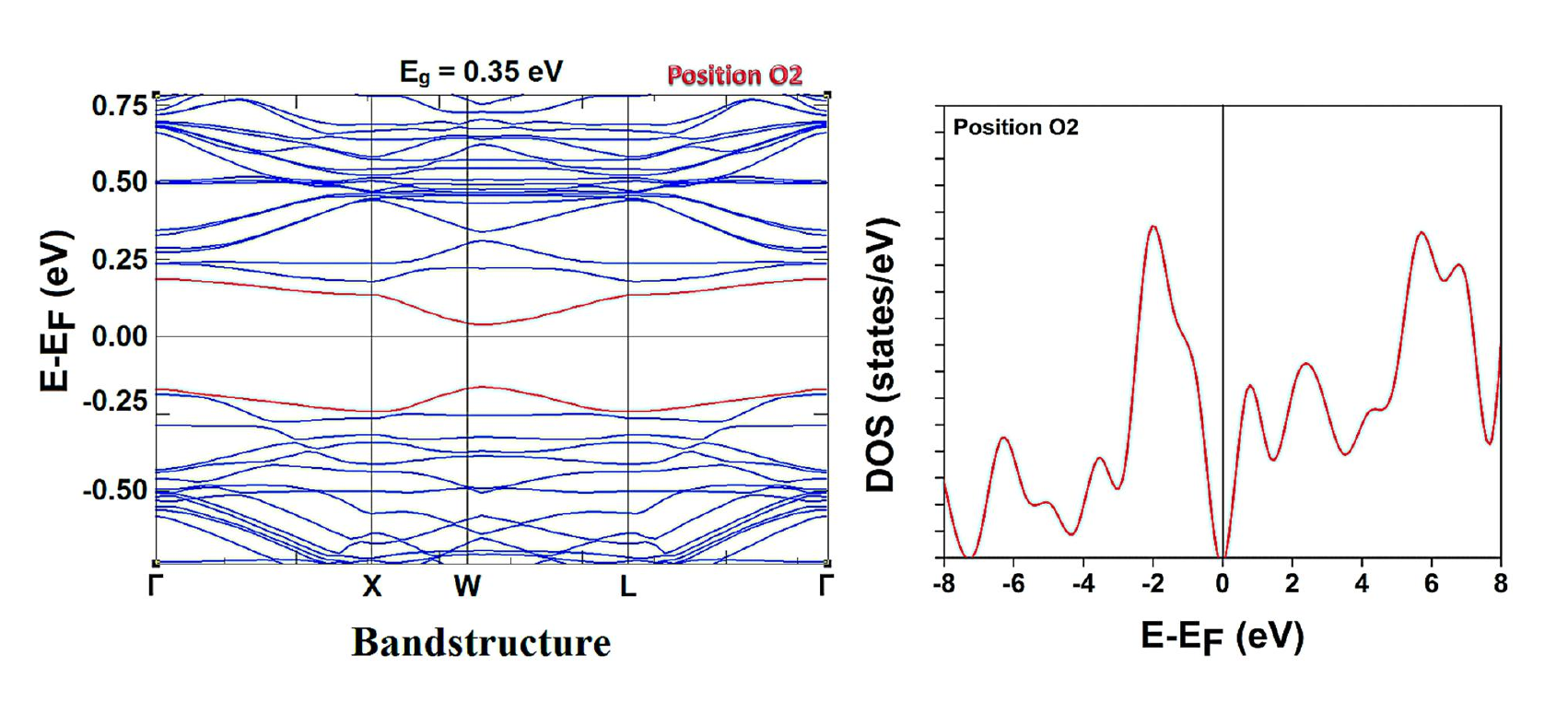}}
\caption{(Colour online) DOS and band structure diagram for position O2.} \label{fig-s15}
\end{figure}

\begin{figure}[!t]
\centerline{\includegraphics[width=0.85\textwidth]{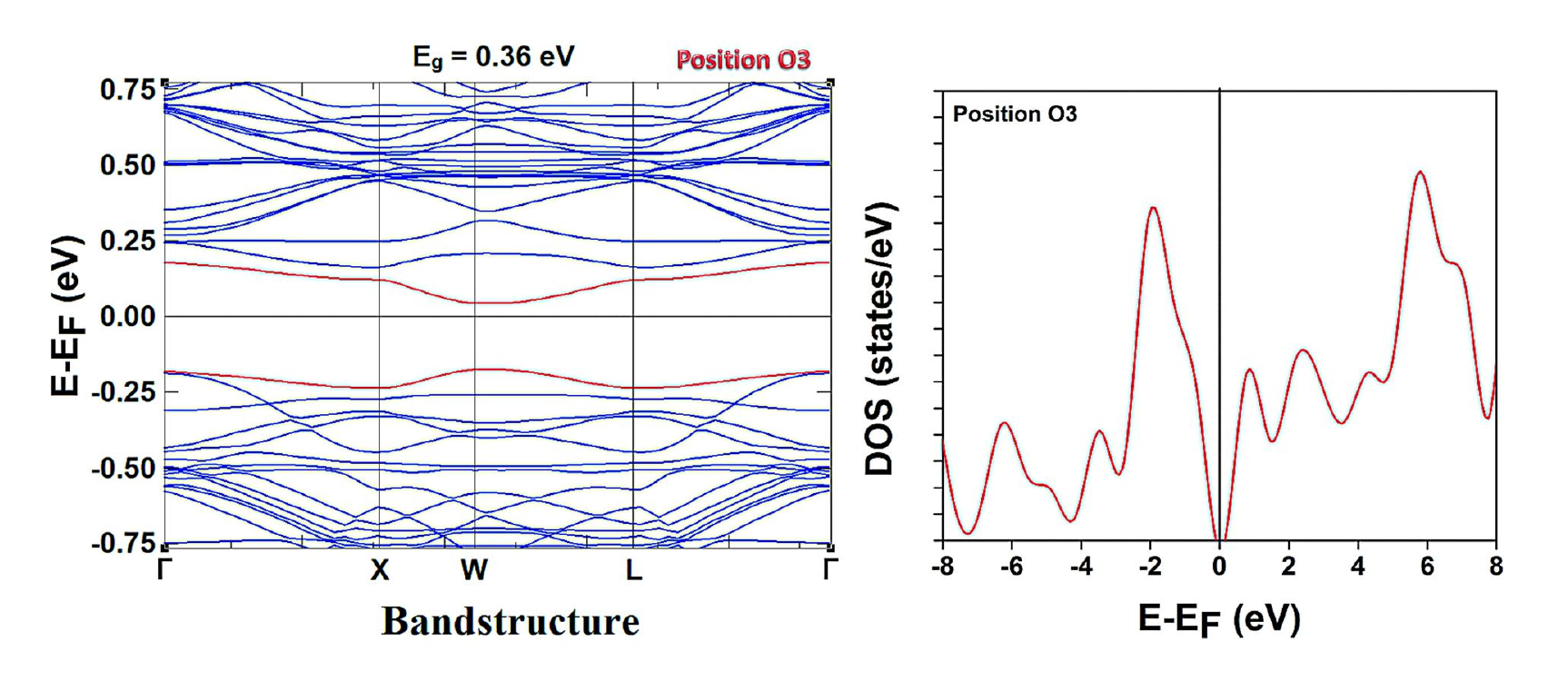}}
\caption{(Colour online) DOS and band structure diagram for position O3.} \label{fig-s16}
\end{figure}

The band structure and the charge density variation in the DOS spectrum clearly reveal that silicene nanosheet can be employed as a substrate to probe the presence of volatiles released from pear fruit \cite{47,48,49,50,51,52}. Figures~\ref{fig-s11}--\ref{fig-s13} exemplifies the band gap structure and DOS-spectrum for positions R1--R3 for the volatiles emitted in the ripened stage and figures~\ref{fig-s14}--\ref{fig-s16} represents the positions O1--O3 for the volatiles emitted in the over-ripened stage of pear fruit. Moreover, the energy band gap gets modulated to 0.41, 0.43 and 0.40 eV, respectively upon adsorption of butyl acetate, butyl butyrate and hexyl acetate on silicene nanosheet. Why does the energy gap of Si-NS get modulated upon the interaction of volatiles? The modulation of the energy gap in silicene takes place due to the interaction of ester molecules that leads to the orbital overlapping with Si atoms in the BSi sheet. Hence, the deviation in the band gap is perceived. Furthermore, the peak maxima in the DOS spectrum for different energy intervals vary, which is advocated due to the orbital overlapping and charge transfer between the pear fruit volatiles and BSi material.

\begin{figure}[!b]
\centerline{\includegraphics[width=0.7\textwidth]{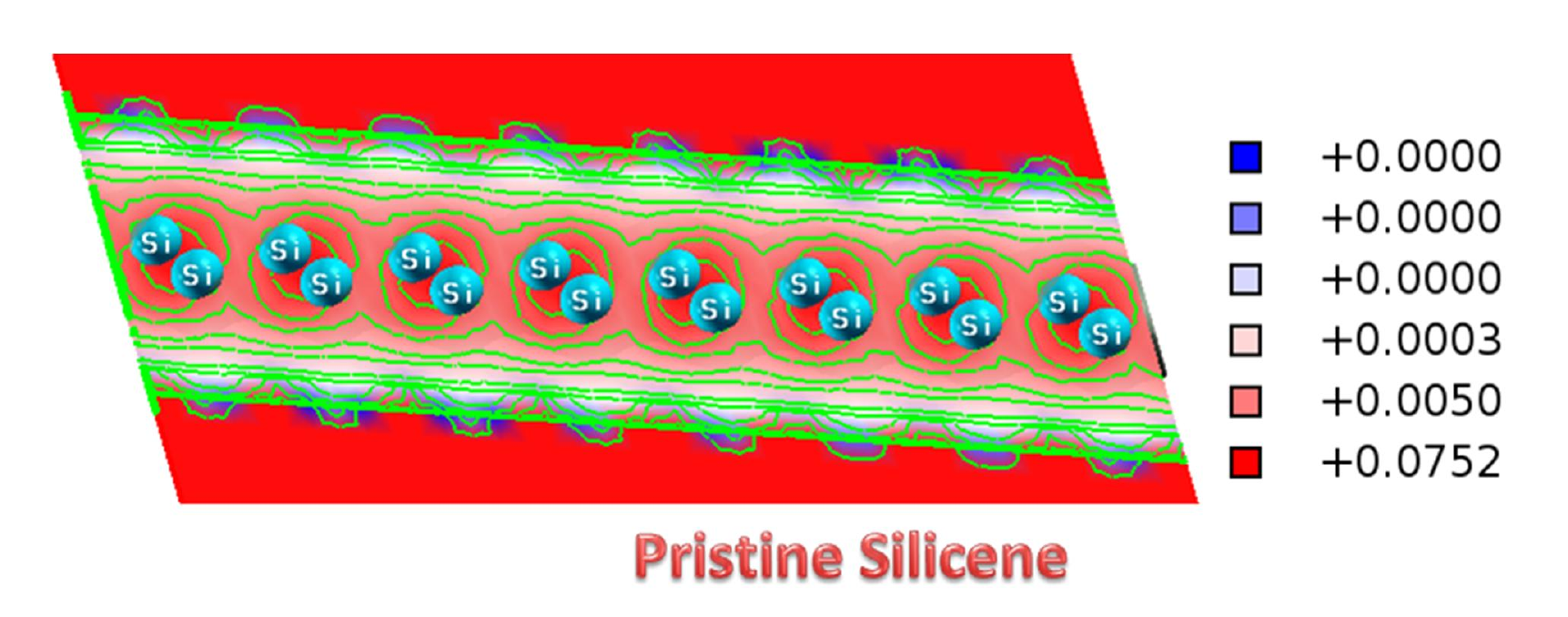}}
\caption{(Colour online) Electron density of pristine silicene nanosheet.} \label{fig-s17}
\end{figure}

\begin{figure}[!t]
\centerline{\includegraphics[width=0.63\textwidth]{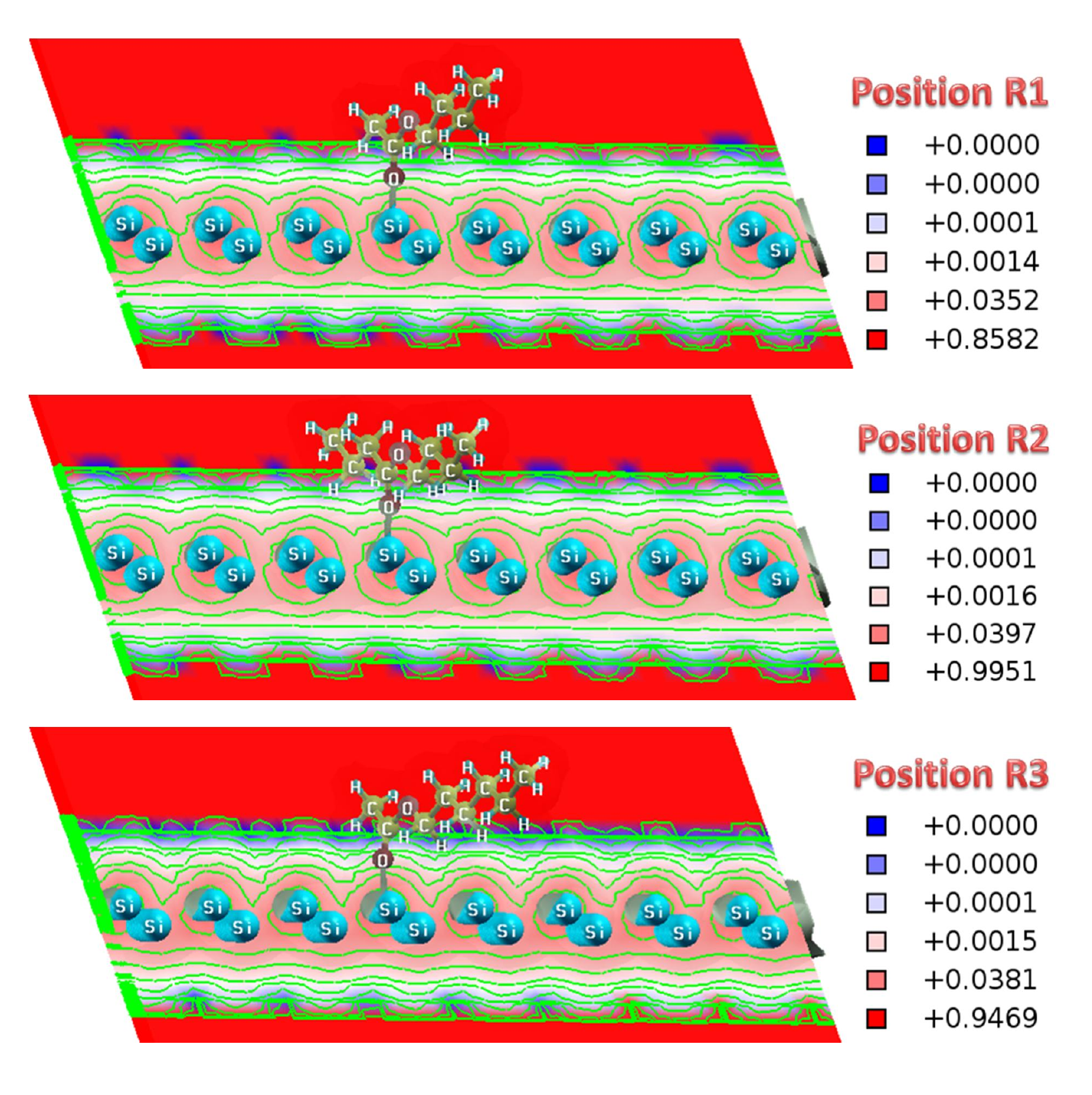}}
\caption{(Colour online) Electron density --- position R1, R2 and R3.} \label{fig-s18}
\end{figure}

\begin{figure}[!t]
\centerline{\includegraphics[width=0.63\textwidth]{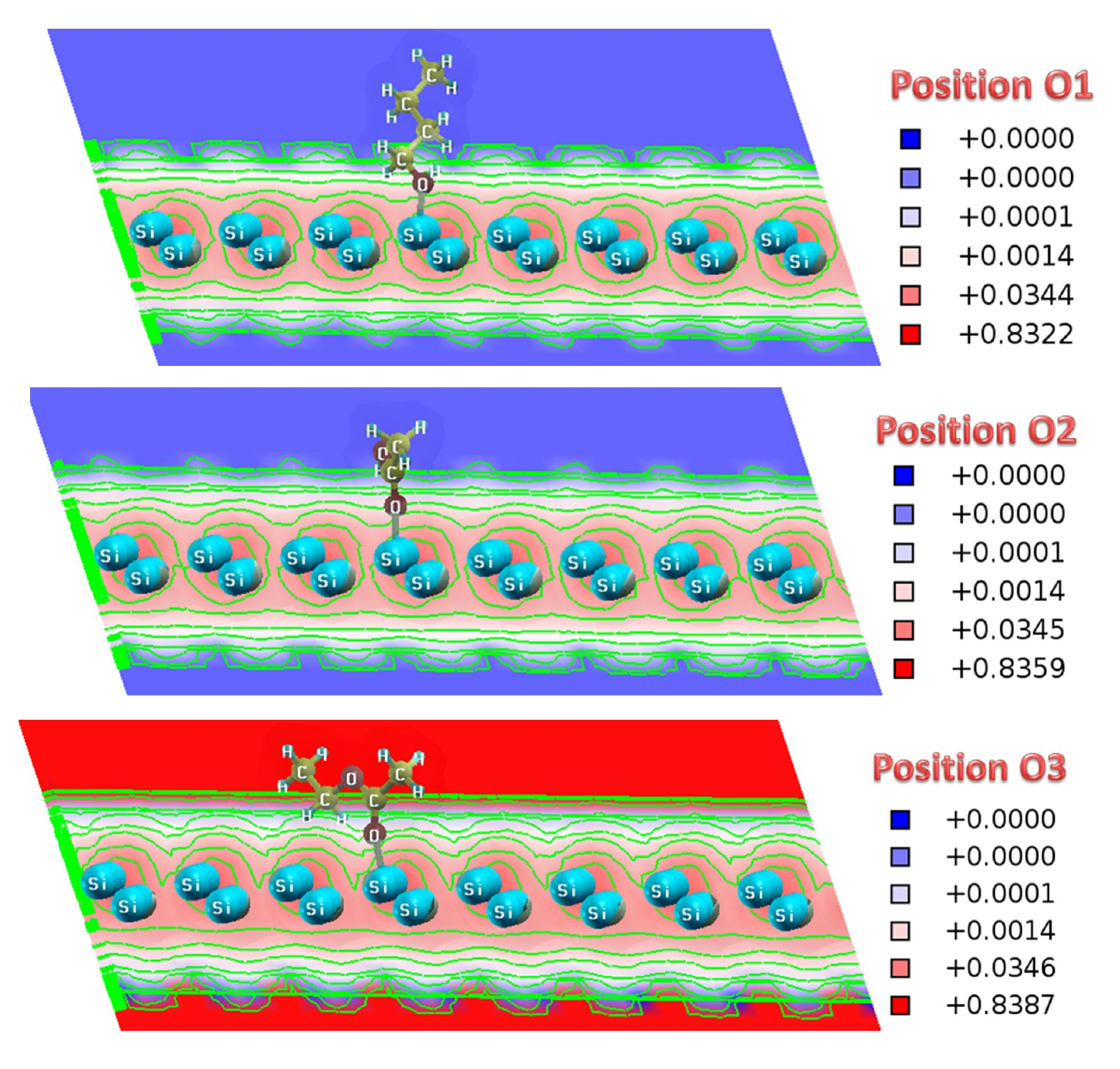}}
\caption{(Colour online) Electron density --- position O1, O2 and O3.} \label{fig-s19}
\end{figure}

Besides, the band gap of silicene nanosheets gets deviated to 0.38 eV, 0.35 eV, and 0.36 eV, respectively. Moreover, the volatiles emitted from pear fruit in the over-ripened stage is slightly lower than that of the volatiles adsorbed in the ripened stage of pear fruit. From the deviation in the band gap structure and DOS spectrum, it is obviously seen that the silicene nanosheet can be used as a two-probe molecular device, which can be used to discriminate the quality of pear fruit. The electron density across the silicene nanosheet and its adsorption over the pear fruit VOCs are depicted in figures~\ref{fig-s17}--\ref{fig-s19}. 

\begin{figure}[!t]
\centerline{\includegraphics[width=0.65\textwidth]{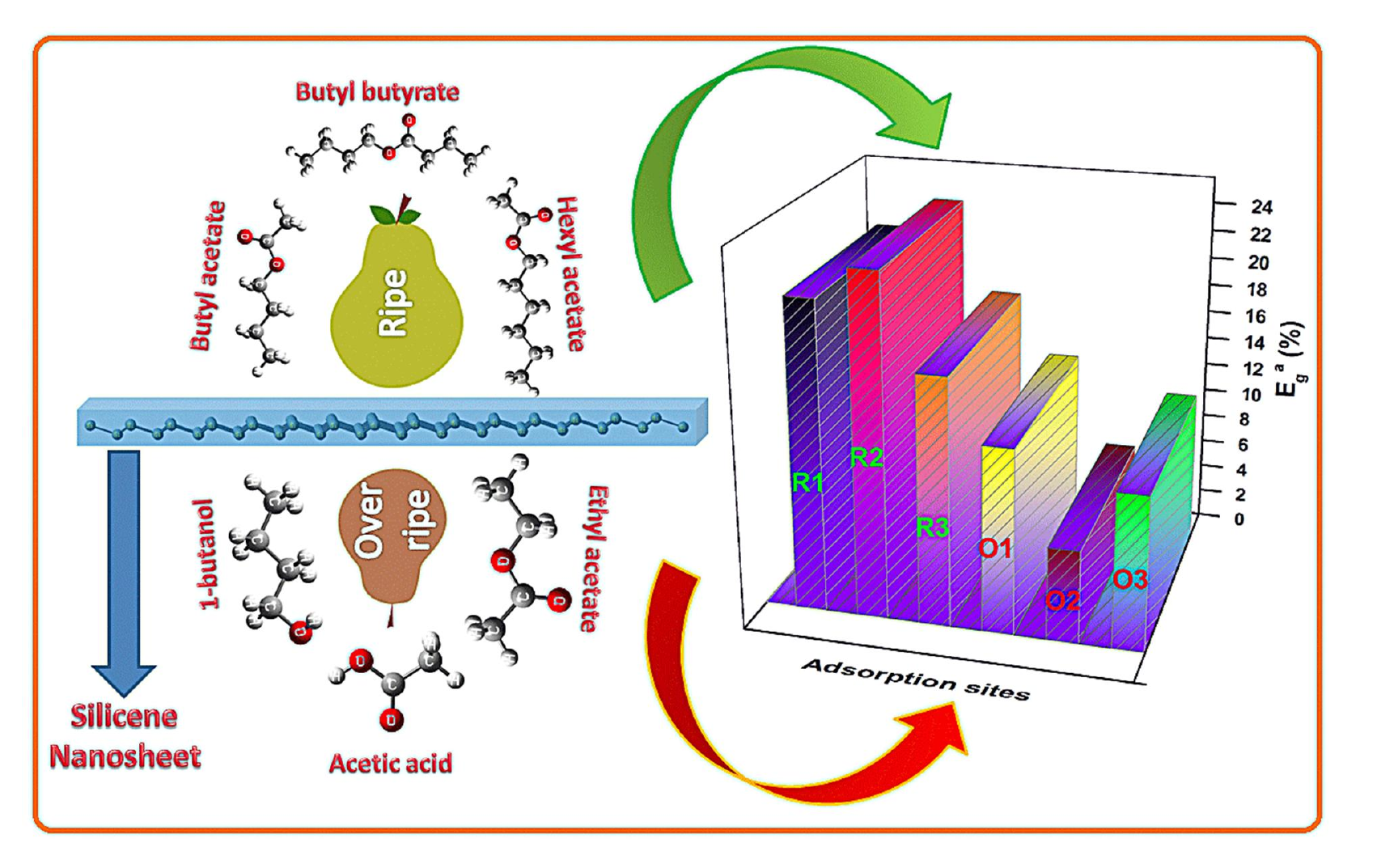}}
\caption{(Colour online) Acumens on the adsorption of VOCs emanated from pear fruit on BSi nanosheet.} \label{fig-s20}
\end{figure}

The charge density is found to change upon the interaction of volatiles emitted from the ripened and over-ripened stage of pear fruit on silicene nanosheet \cite{53,54,55}. Thus, it is evident that the charge density varies upon adsorption of VOCs on the BSi sheet. Figure~\ref{fig-s20} gives clear insights on the adsorption of VOCs released from pear fruit in ripened and over-ripened stage on to the silicene nanosheet.

\section {Concluding remarks}
Using DFT method, we deliberated adsorption property of VOCs of pear fruit in ripened and over-ripened stages on to the silicene nanosheet. The adsorption of VOCs of pear fruit on 2D monolayer silicene base material is studied using adsorption energy, energy band gap, and charge transfer. The VOCs of pear fruit, namely butyl acetate, butyl butyrate, hexyl acetate, 1-butanol, acetic acid, and ethyl acetate is physisorbed on silicene nanosheet. The interaction of VOCs of pear fruit is in the order of hexyl acetate $\rightarrow$ butyl acetate $\rightarrow$ butyl butyrate for the ripened stage whereas in the over-ripened stage the adsorption sequence is in the order, acetic acid $\rightarrow$ ethyl acetate $\rightarrow$ 1-butanol. Moreover, the drastic dissimilarity in the band gap is observed for VOCs of pear fruit such as acetic acid, ethyl acetate, and 1-butanol found in the over-ripened stage upon adsorption on silicene nanosheet. Thus, we reason out that a simple Si-NS material can be utilized as a two-probe molecular device to study the adsorption property of VOCs present in pear fruit. The findings suggest that Si-NS material can be utilized as one of the prominent base materials for quality checking of pear fruit.

\section {Acknowledgements}

The authors wish to express their sincere thanks to Nano Mission Council (No.SR/NM/NS- \linebreak 1011/2017(G)) Department of Science \& Technology, India for the financial support.

\ukrainianpart
\title{Силіценовий нанолист  для розпізнавання  якості груш, що використовує адсорбцію летких речовин --- застосування~DFT  }%
\author{Р. Кірсі Бхавадхарані\refaddr{label1}, В. Нагараджан\refaddr{label2}, Р. Чандірамулі\refaddr{label2} }
\addresses{
\addr{label1} 
Школа хімічної біотехнології, унiверситет SASTRA, Тiрумалайсамудрам, Танджавур --- 613 401, Iндiя 
\addr{label2} Школа електротехнiки та електронiки, унiверситет SASTRA, Тiрумалайсамудрам, Танджавур --- 613 401, Iндiя 
}

\makeukrtitle

\begin{abstract}
 Використовуючи метод теорії функціоналу густини, ми представляємо взаємодію між силіценовим  нанолистом
 (Si-NS) і леткими органічними сполуками, які   вивільняються з груш  (Pyrus communis) на стадіях стиглості  і перестиглості. Геометрична стійкість  Si-NS вивчається зі структури фононного спектру. Далі, електронна властивість  Si-NS вивчається із енергетичної структури забороненої зони, і знайдено, що енергія забороненої зони дорівнює  0.46~еВ і має напівпровідникову природу. Результати показують, що адсорбція летких речовин, які вивільняються з груш на силіценовий нанолист, є в такому порядку: гексил ацетат $\rightarrow$ бутил ацетат $\rightarrow$ бутил  бутират на стадії стиглості, в той час як на стадії перестиглості відмічається така послідовність адсорбції: оцтова кислота $\rightarrow$ етил ацетат $\rightarrow$ 1-бутанол. Характеристику адсорбції летких речовин, що вивільняються з груш на силіценовий нанолист, задокументовано з енергією адсорбції, зміною середньої енергії забороненої зони і переносом заряду Бадера. Окрім того, адсорбція летких органічних сполук на силіценовий нанолист також вивчається, використовуючи енергетичну зонну структуру, електронну густину разом з адсорбційними вузлами і спектр густини станів. Також, результати показали, що силіценовий нанолист може бути використано для розпізнавання якості груш.

\keywords силіцен, нанолист, груша, адсорбція, леткі речовини, заборонена зона %

\end{abstract}

\end{document}